\documentstyle[aps,twocolumn,tighten,epsfig]{revtex}
\begin{document}

\draft
\title{ DENSITY OF KINKS JUST AFTER A QUENCH IN AN OVERDAMPED SYSTEM }

\author{Jacek Dziarmaga\thanks{E-mail: {\tt ufjacekd@thp1.if.uj.edu.pl}}}
\address{ Institute of Physics, Jagiellonian University,
          Reymonta 4, 30-059 Krak\'ow, Poland}
\date{February 19, 1998}
\maketitle
\tighten

\begin{abstract}
{\bf A quench in an overdamped
one dimensional $\phi^4$ model is studied by analytical and numerical
methods. For
an infinite system or a finite system with free boundary conditions,
the density of kinks after the transition is proportional to the eighth
root of the rate of the quench. For a system with periodic boundary
conditions, it is proportional to the fourth root of the rate. The critical
exponent predicted in Zurek scenario is put in question. }
\end{abstract}
\vspace*{1cm}

   Topological defects play a prominent role in many condensed matter systems,
see e.g.\cite{davydov} for a review. It was also suggested that they
were an important ingredient of the early universe \cite{vilenkin}.
Topological defects can be generated in large numbers during
a second order phase transition. The dynamics of such a transition
to a symmetry broken phase has been an object of much recent attention
because of its importance in the cosmological context \cite{cosmos} and in
condensed matter systems \cite{condmat}.

   An early estimate of defect density after a quench was given by Kibble
\cite{kibble}. It his theory, the speed of light is a dominant factor
which determines the size of correlated domains. An alternative scenario
was put forward by Zurek \cite{zurek}, who emphasized the importance of
the nonequilibrium dynamics of the order parameter. Recent experiments point to
the latter theory. The principal prediction of Ref.\cite{zurek} is the
scaling law. He is able to predict by very general arguments an exponent
in a power law dependence of the density of topological defects on a time
of the quench in a variety of systems. The numerical experiment \cite{lz}
seems to confirm this prediction. 
  
   It is normally assumed that close to the phase transition
the dynamics of the system is overdamped. We follow this line of argument
to its logical limit and concentrate from the very beginning on a purely
diffusive dynamics. The model we study is the $\phi^4$ theory in
one spatial dimension. Although there is no proper phase transition in such
a one dimensional model there is a long range order in its symmetry broken
phase. The real field $\phi$ falls into one of the two minima of the
Ginzburg-Landau potential at $\phi=+1$ or $\phi=-1$. The domain
walls separating these two different vacua are the topological solitons
or the kinks. We develop a novel method of how to predict the density
of kinks after a quench. It turns out that the density is proportional
to the eighth root of the quench rate in contradiction with Zurek's
prediction of the fourth root. We point to an implicit assumption in Zurek 
scenario, which lays at the bottom of this discrepancy.

   Let us consider a one dimensional overdamped $\phi^4$ model

\begin{equation}\label{model}
\dot{\phi}(t,x) = \phi''(t,x) - 2 a(t) \phi(t,x) - 2 \phi^{3}(t,x) +
                  \eta(t,x)    \;\;,
\end{equation}
where $\dot{}\equiv\partial{t}$, $ '\equiv\partial_{x}$. $\phi(t,x)$ is
a real order parameter. $\eta(t,x)$ is a Gaussian noise of temperature
$T$ with correlations

\begin{eqnarray}\label{correlations}
&& <\eta(t,x)>=0 \;\;, \nonumber\\
&& <\eta(t_1,x_1)\eta(t_2,x_2)>= 2 T \delta(t_1-t_2)\delta(x_1-x_2) \;\;.
\end{eqnarray}
The coefficient $a(t)$ is time dependent. We consider a linear quench

\begin{equation}\label{a(t)}
a(t)=
\left\{
\begin{array}{ll}
A                   & \mbox{ , if $\; t<0$ }                   \\
A(1-\frac{t}{\tau}) & \mbox{ , if $\; 0<t<\tau\frac{A+1}{A}$ } \\
-1                  & \mbox{ , if $\; \tau\frac{A+1}{A}<t$ }
\end{array}
\right. 
\end{equation}
Before the quench, for $t<0$, the system is in a symmetric phase $(A>0)$,
during the quench, at $t=\tau$, it undergoes a transition from the symmetric
phase $a(t<\tau)>0$ to a broken symmetry phase $a(t>\tau)<0$. Finally it
settles down at $a(t)=-1$.

  As long as the system is still in the symmetric phase ($t<\tau$),
the field $\phi$ can be regarded as a small fluctuation around its
symmetric ground state $\phi(t,x)=0$. It is justified to neglect
the cubic term on the RHS of Eq.(\ref{model}).
The field can be written as a Fourier transform,

\begin{equation}\label{fourier}
\phi(t,x)=\int_{-\infty}^{+\infty}dk\; \tilde{\phi}(t,k) e^{ikx} \;\;.
\end{equation}
The Fourier transform of linearized Eq.(\ref{model}) is

\begin{equation}\label{modelfourier}
\dot{\tilde{\phi}}(t,k)=-[2 a(t)+k^2]\tilde{\phi}(t,k)+\tilde{\eta}(t,k)\;\;.
\end{equation}
The Fourier transformed noises have correlations

\begin{eqnarray}\label{correlationsfourier}
&& <\tilde{\eta}(t,k)>=0  \;\;,\nonumber \\
&& <\tilde{\eta}^{*}(t_1,k_1)\tilde{\eta}(t_2,k_2)>=
   \frac{T}{\pi}\delta(t_1-t_2)\delta(k_1-k_2) \;\;.
\end{eqnarray}
The formal solution of Eq.(\ref{modelfourier}) at the time $\tau$ is

\begin{equation}
\tilde{\phi}(\tau,k)=\int_{-\infty}^{\tau}dt_1\;
  \exp\{ -\int_{t_1}^{\tau} dt_2 \; [2 a(t_2) + k^2] \}\;
  \tilde{\eta}(t,k) \;\;.
\end{equation}
This solution, the correlations (\ref{correlationsfourier}) and
the explicit form of $a(t)$, see Eq.(\ref{a(t)}), give the correlation
function at $t=\tau$

\begin{eqnarray}\label{phiphi}
&& <\tilde{\phi}^{*}(t_1,k_1)\tilde{\phi}(t_2,k_2)>= 
   \frac{T}{\pi}\delta(k_1-k_2)\;G(\tau,A,k)         \nonumber\\
&& G(\tau,A,k)=
   \frac{e^{-2 k^2 \tau}}{4A+2 k^2}+                 \nonumber\\
&&\;\;\;\; \sqrt{\frac{\pi\tau}{8A}}
  [Erf(\sqrt{\frac{\tau}{2A}}\;(2A+k^2))-Erf(\sqrt{\frac{\tau}{2A}\;k^2})] \;\;.
\end{eqnarray}
The linearized approximation can not be used in the broken symmetry
phase. Nevertheless we would like to use the correlations at $t=\tau$
to predict the density of topological defects after the quench.
We have to make contact between the formalism developed for the
symmetric phase and the one which is relevant to the symmetry broken
phase.

  $a(t)=-1$ in the symmetry broken phase. If the noise is neglected,
then Eq.(\ref{model}) has a static kink solution

\begin{equation}\label{kink}
\phi(t,x)=\tanh(x-z)\equiv K(x-z)
\end{equation}
with an arbitrary kink position $z$. The field which is going to relax
to this equilibrium configuration (\ref{kink}) can be expressed
as a sum of the kink solution and of eigenfunctions $u_{\alpha}$ of the
fluctuation operator around the kink numbered by a discrete/continuous index
$\alpha$,

\begin{equation}\label{alpha}
\phi(t,x)=K(x-z)+\sum_{\alpha} \Phi_{\alpha}(t) u_{\alpha}(x-z) \;\;.
\end{equation}
The eigenstates have positive eigenvalues and their amplitudes
$\Phi_{\alpha}(t)$ decay exponentially with time. The
"kink transform" of the field (\ref{alpha})

\begin{equation}\label{kinktransform}
I(t,x)=\int_{-\infty}^{+\infty} dy\; K'(y) \phi(t,x+y)
\end{equation}
vanishes for $x=z$ because $K'(x)$ is orthogonal to $K(x)$ and to
$u_{\alpha}(x)$'s. The zero of the kink transform of the initial
field is the equilibrium position of the kink. The same transform
(\ref{kinktransform}) can be used to predict positions of antikinks.
This method has been introduced and extensively tested numerically
in \cite{pla}. To predict the positions of kinks after the quench
one must find the kink transform $I(\tau,x)$ of the initial field
$\phi(\tau,x)$ and locate the zeros of $I(\tau,x)$.

  We do not need any detailed knowledge about the actual positions
of kinks. All we need to know is their average density. The
density of kinks $n$ can be expressed by the density of zeros of the
kink transform $I(\tau,x)$

\begin{equation}
n=\frac{<N>}{2L}=
\frac{1}{2L}\int_{-L}^{L} dx\; <|I'(\tau,x)|\;\delta[I(\tau,x)]> \;\;.
\end{equation}
$L$ is the length of the considered interval. Due to
translational invariance

\begin{equation}
n=<|I'(\tau,0)|\;\delta[I(\tau,0)]> \;\;.
\end{equation}
Thanks to the same symmetry $<I(\tau,x)I'(\tau,x)>=0$ for any $x$,
so that

\begin{equation}
n=<|I'(\tau,0)|>\;<\delta[I(\tau,0)]>.
\end{equation}
The identity $|I'|=I'\;Sign(I')$ and the Fourier
transforms of the $\delta$ and $Sign$ functions help to derive

\begin{equation}\label{n}
n=\;\frac{\pi}{2}\;\sqrt{\frac{<I'^2(\tau,0)>}{<I^2(\tau,0)>}}.
\end{equation}
The higher is the ratio of the variation of $I$ to its average magnitude,
the higher is the density of kinks. The higher is this ratio, the
more likely is the function $I$ to cross zero.

One can express the kink transform with the Fourier transform of
$\phi(\tau,x)$ (\ref{fourier}) as

\begin{eqnarray}
&& I(\tau,0)=\int_{-\infty}^{+\infty}dk\;U(k)\;\tilde{\phi}(\tau,k)
                                                      \;\;,\nonumber\\
&& I'(\tau,0)=i\int_{-\infty}^{+\infty}dk\;k\;U(k)\;\tilde{\phi}(\tau,k)
                                                      \;\;,\nonumber\\
&& U(k)=\frac{\pi k}{\sinh(\pi k/2)} \;\;.
\end{eqnarray}
$U(k)$ is a Fourier transform of $K'(x)$. The $I-I$ correlations
in Eq.(\ref{n}) can be expressed by the correlations (\ref{phiphi}),
so that finally we obtain

\begin{equation}\label{density}
n=\;\frac{\pi}{2}\;
  \sqrt{\frac{ \int dk\;k^2 U^2(k) G(\tau,A,k) }
             { \int dk\;    U^2(k) G(\tau,A,k) }} \;\;.
\end{equation}
The kink size provides an ultraviolet cut-off through the function $U^2(k)$.
Without any UV cut-off the expression (\ref{density}) would be infinite
for any $A$ and $\tau$. In the symmetric phase the noise driven order
parameter $\phi(t,x)$ has an infinite number of zeros.

  To extract the scaling law, we must find the asymptote of $n$
for large $\tau$. The asymptote of the correlation function $G$ is

\begin{equation}\label{G}
G(\tau,A,k)\approx
           \sqrt{\frac{\tau}{2A}}
           e^{2A\tau+\frac{k^4 \tau}{2A}}
           \int_{\sqrt{\frac{\tau}{2A}}k^2}^{+\infty} dx\;e^{-x^2} \;\;.
\end{equation}
We can make further approximation under the integral
in the numerator of (\ref{density})

\begin{eqnarray}
&& \int_{-\infty}^{+\infty} dk\;k^2\; U^2(k)\; G(\tau,A,k)\approx \nonumber\\
&& \frac{1}{2} e^{2A\tau}\int_{-\infty}^{+\infty} dk\;U^2(k)=  
   \frac{4\pi}{3} e^{2A\tau} \;\;.
\end{eqnarray}
We introduce $p=k\;(\frac{\tau}{2A})^{1/4}$ in the denominator integral
to obtain the asymptote

\begin{eqnarray}
&& \int_{-\infty}^{+\infty} dk\; U^2(k) G(\tau,A,k)\approx \nonumber\\
&& (\frac{\tau}{2A})^{\frac{1}{4}} e^{2A\tau}
   \int_{-\infty}^{+\infty} dp\;
     U^2[\frac{p}{  (\frac{\tau}{2A})^{\frac{1}{4}}  }]\;
     e^{p^4}\int_{p^2}^{+\infty} dx\; e^{-x^2} \approx     \nonumber\\
&& (\frac{\tau}{2A})^{\frac{1}{4}} e^{2A\tau}\; U^2(0)
   \int_{-\infty}^{+\infty} dp\;
   e^{p^4}\int_{p^2}^{+\infty} dx\; e^{-x^2}\;.
\end{eqnarray}
After the numerical coefficients are worked out one obtains

\begin{equation}\label{nfree}
n\approx 1.16\;(\frac{A}{\tau})^{\frac{1}{8}} \;\;.
\end{equation}
The density of kinks is proportional to the eighth root of the
transition rate. The asymptote is achieved for $\tau>>2A$. The
critical exponent does not depend on the actual choice of the UV cut-off
provided by the function $U^2(k)$. In a lattice theory, regularized
by its lattice constant, one would get the $1/8$ critical exponent
for the density of zeros of the order parameter $\phi(\tau,x)$ itself.

  The last prediction must be contrasted with a corresponding result
for a model with periodic boundary conditions. The periodicity
$\phi(t,L)=\phi(t,0)$ implies that momentum is quantized
$k_m=(\frac{2\pi}{L})m$ with integer $m$. The integrals over $k$ in
(\ref{density}) are replaced by summations over $m$ and all appearances
of $k$ are replaced by $k_m$. The denominator sum is dominated for large
$\tau$ by the $m=0$ mode contribution,

\begin{eqnarray}
&& \sum_{m=-\infty}^{+\infty} U^2(k_m) G(\tau,A,k_m) \approx \nonumber\\
&& \sum_{m=-\infty}^{+\infty} U^2(k_m)
     \sqrt{\frac{\tau}{2A}}\;
     e^{2A\tau+\frac{k_m^4\tau}{2A}}
     \int_{\sqrt{\frac{\tau}{2A}}k_m^2}^{+\infty} dx\;e^{-x^2}\approx\nonumber\\
&& \sqrt{\frac{\tau}{2A}}\;
   e^{2A\tau}\;
   U^2(0)\int_{0}^{+\infty} e^{-x^2}  \;\;.
\end{eqnarray}
The numerator sum is approximated in a similar way as the numerator
integral for the infinite system,

\begin{eqnarray}
&& \sum_{m=-\infty}^{+\infty} k_m^2\; U^2(k_m)\; G(\tau,A,k_m)\approx\nonumber\\
&& \sum_{m=-\infty}^{+\infty} k_m^2\; U^2(k_m)\;
     \sqrt{\frac{\tau}{2A}}\;
     e^{2A\tau+\frac{k_m^4\tau}{2A}}
     \int_{\sqrt{\frac{\tau}{2A}}k_m^2}^{+\infty} dx\;e^{-x^2}\approx\nonumber\\
&& \frac{1}{2}e^{2A\tau}\sum^{+\infty}_{\stackrel{m=-\infty}{m\neq 0}} U^2(k_m)
\end{eqnarray}
Putting these two asymptotes together we get

\begin{equation}
n\approx \; 
\sqrt{ \frac{ \pi^{\frac{3}{2}}\sum_{m} U^2(k_m) }{ 8 } }\;\;
(\frac{A}{\tau})^{\frac{1}{4}} \;\;.
\end{equation}
If the periodic boundary conditions are imposed, the density of
kinks is proportional to the fourth root of the transition rate.
In fact this conclusion is the same for any fixed boundary conditions
which quantize $k$. The asymptote is achieved for
$\sqrt{\frac{\tau}{2A}}k_1^2=\sqrt{\frac{\tau}{2A}}\frac{(2\pi)^2}{L^2}>>1$.
We expect for a large but finite system, that the density scales like
the eighth root of the quench rate for moderately large $\tau$
but for very large $\tau$ it scales like the fourth root.
Transition between these two regimes takes place at a $\tau$ which
grows like the fourth power of the system size $L$. The thermodynamic
limit and the large quench time limit do not commute. Let us note in passing
that a similar dependence of critical exponent on boundary conditions
takes place in the diffusion of overdamped kinks in the sine-Gordon
chain \cite{buttiker}. 

   The Zurek scenario as described in \cite{zurek} does not refer to boundary
conditions. Let us analyse our derivation from the point of view of
Ref.\cite{zurek}. The relaxation time of the system depends on time like
$t_r=\frac{1}{2a(t)}$. The system ceases to follow the changes in $a(t)$
at the instant $\hat{t}$, when its relaxation time becomes greater than
the time left to the transition at $t=\tau$,
$\;(\tau-\hat{t})=\frac{1}{2a(\hat{t})}$. For the linear quench (\ref{a(t)})
this takes place when $(\tau-\hat{t})=\sqrt{\frac{\tau}{2A}}$.
At this instant of time the correlation length is
$\hat{\xi}=(\frac{\tau}{2A})^{1/4}$. The density of kinks after the
transition is inversely proportional to this "frozen" correlation length,
so $n\sim (\frac{A}{\tau})^{1/4}$.

   The weak point of this argument is that $t_r=\frac{1}{2a(t)}$
is the relaxation time but only for the $k=0$ mode. Modes with nonzero
momenta $k>0$ have relaxation times $t_r(k)=\frac{1}{2a(t)+k^2}<t_r(0)$
and they are frozen later than the $k=0$ mode. As we can see in
Eq.(\ref{n}), the density of nucleated
kinks is related to the ratio of the magnitude of short wavelength modes
to the magnitude of long wavelength modes. Contrary to Zurek scenario,
which implicitly
assumes that all the modes are frozen at the same instant of time,
the $k>0$ modes can still grow for a while even after the $k=0$ mode
was already frozen. Thus the density of kinks after quench is greater than
that given by Zurek estimate. It is a matter of detailed calculations presented
above to find out that the actual critical exponent is $1/8$ instead
of $1/4$.

   It is the point to discuss the numerical experiment of Laguna and
Zurek \cite{lz}. Those authors discuss the model (\ref{model}) 
with $\ddot{\phi}$ added on the LHS. Their numerical 
experiment on a lattice with periodic boundary conditions gives a
crititical exponent of $1/4$. We could naively
argue that this is due to the periodic boundary conditions and in agreement
with our theory. This is, however, not the case. The lattice \cite{lz} 
is large enough to render the boundary conditions irrelevant. Hence
according to our theory one would expect the $1/8$ scaling.
But the point is that neither Zurek estimate nor our calculations apply
to this system. The damping coefficient $\eta=1$ used in actual
simulations \cite{lz}
(in distinction to the theoretical discussion \cite{lz}) is in the
underdamped range. At the beginning of the quench all the field
oscillators with different momenta $k$ are underdamped. The $k=0$ would
be overdamped, if $\eta>\sqrt{2}$ in units of Ref.\cite{lz}.
The short wavelength modes would require even higher $\eta$.
At the beginning of the quench the second order time
derivative dominates over the first order derivative.
Without any reliable tool at hand we can only make a very superficial
remark. Replacement of $\dot{\phi}$ by $\ddot{\phi}$ should give
the replacement of $\tau$ by $\tau^2$ in any scaling behavior. This
replacement in our formula (\ref{nfree}) leads to the critical
exponent $1/4$ in agreement with the numerics of Ref.\cite{lz}.

  We performed a numerical simulation of the quench (\ref{a(t)}) in
the model (\ref{model}). The total extension of our lattice was
$20$ units, the lattice constant $0.1$ and the temperature $T=10^{-8}$.
With these parameters the periodic lattice system was far in the thermodynamic
regime, the asymptote with the $1/4$ scaling is expected to be
achieved for $\tau>> 10^4$. 
The parameter $A$ in Eq.(\ref{a(t)}) was chosen as $A=100$. We have 
measured the density of zeros of the function $\phi(\tau,x)$ for 
$\tau=0.001 \ldots 3.2$ units of time. The simulation was repeated
20 times for each $\tau$. The figure shows the double logarithmic
plot of the density of zeros $n_{z}$ as a function of the quench
time. The slope is $-0.110\stackrel{+}{-}0.005$, which is consistent
with our prediction of $-\frac{1}{8}=-0.125$.

\subsection*{Conclusion.}

The critical exponent observed in an experiment on a macroscopic system in
its overdamped regime should be $1/8$. One can observe the $1/4$
scaling in a numerical "experiment" on a small lattice.

\acknowledgements
I would like to thank Markus B\"uttiker for sending me a reprint of 
Ref.\cite{buttiker}.

\centerline{\epsfbox{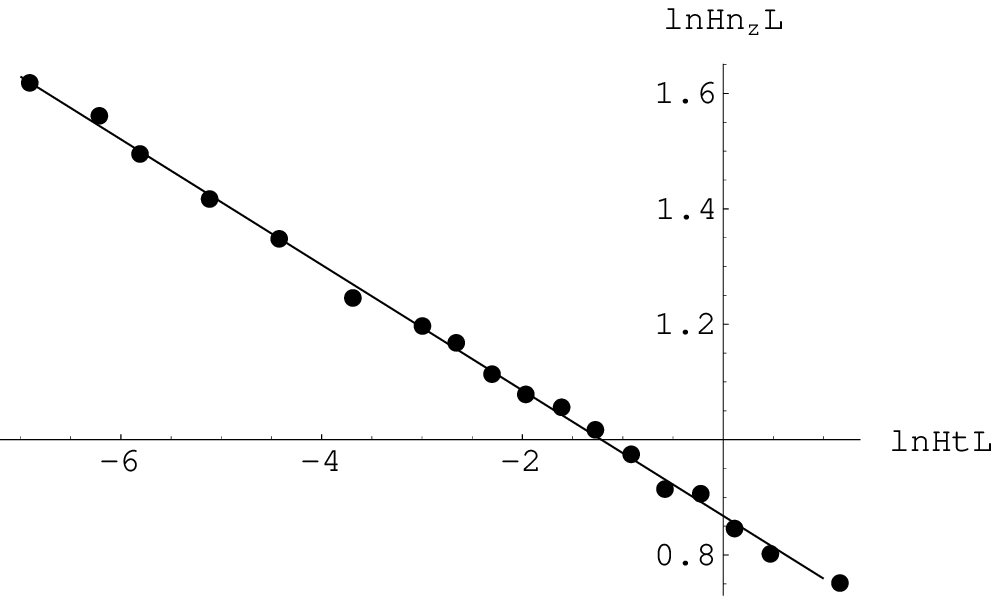}}

{\bf Figure caption.}
Logarithm of the density of zeros of the function $\phi(\tau,x)$
as a function of the logarithm of the quench time $\tau$. The slope
of the fitted line is $-0.110\stackrel{+}{-}0.005$, which
is consistent with the prediction of $-\frac{1}{8}=-0.125$.

\end{document}